\documentstyle[12pt]{article}\textheight 230mm\textwidth 150mm
            \newcommand{\be}{\begin{eqnarray}}
            \newcommand{\ee}{\end{eqnarray}}
           \newcommand{\eel}[1]{\label{#1}\end{eqnarray}}
\newcommand{\e}[1]{\label{e:#1}\end{eqnarray}}
     \newcommand{\eg}{{\em e.g.\ }}
            \newcommand{\ie}{{\em i.e.\ }}
            \newcommand{\ga}{{\gamma}}
 
            \newcommand{\la}{{\lambda}}
            
            \newcommand{\del}{{\delta}}

\newcommand{\bp}{\bar{p}}
\newcommand{\bla}{\bar{\lambda}}

\newcommand{\bmu}{\bar{\mu}}
   
\newcommand{\dx}{\dot{x}}
\newcommand{\da}{\dot{a}}

 \newcommand{\ca}{{{\cal C}}}
\newcommand{\bac}{{\bar{\cal C}}}
\newcommand{\cH}{{{\cal H}}}

            \newcommand{\pet}{{\cal P}}
\newcommand{\bpet}{\bar{\cal P}}

            \newcommand{\beq}{\begin{quote}}
            \newcommand{\eq}{\end{quote}}
            \newcommand{\Om}{\Omega}
            \newcommand{\al}{\alpha}
            \newcommand{\ben}{\begin{enumerate}}
            \newcommand{\een}{\end{enumerate}}
            \newcommand{\bit}{\begin{itemize}}
            \newcommand{\ei}{\end{itemize}}
    	\newcommand{\nn}{\nonumber}
            \newcommand{\r}[1]{(\ref{e:#1})}
            \newcommand{\edfl}[1]{\Label{#1}\end{df}}

\newcommand{\hb}{{\cal i}}

\newcommand{\ve}{{\varepsilon}}

\newcommand{\dagg}{^{\dag}}

\newcommand{\dif}{{\partial}}
\newcommand{\half}{\frac{1}{2}}

\begin{document}
\begin{titlepage}

\vspace*{5 mm}
\vspace*{20mm}
\begin{center}
{\LARGE\bf
Projection operator approach to}\end{center}\begin{center}{\LARGE\bf
general constrained systems}\end{center}
\vspace*{3 mm}
\begin{center}
\vspace*{3 mm}

\begin{center}Igor A. Batalin\footnote{On leave of absence from
P.N.Lebedev Physical Institute, 117924  Moscow, Russia\\E-mail:
batalin@td.lpi.ac.ru.}, Simon L. Lyakhovich\footnote{On leave of absence from
Department of Theoretical Physics, Tomsk State University, 634050 Tomsk,
Russia. \
E-mail: sll@phys.tsu.ru.} and Robert
Marnelius\footnote{E-mail: tferm@fy.chalmers.se}
 \\ \vspace*{7 mm} {\sl
Department of Theoretical Physics\\ Chalmers University of Technology\\
G\"{o}teborg University\\
S-412 96  G\"{o}teborg, Sweden}\end{center}
\vspace*{25 mm}
\begin{abstract}
We propose a new  BRST-like quantization procedure which is applicable to
dynamical systems containing both first and second class constraints. It
requires no
explicit separation into first and second class constraints and therefore no
conversion of second class constraints is needed. The basic ingredient is
instead an
invariant projection operator which projects out the maximal subset of
constraints in
involution. The hope is that the method will enable a covariant quantization of
models for which there is no covariant separation into first and second class
constraints. An example of this type is given.
\end{abstract}\end{center}\end{titlepage}

\setcounter{page}{1}
\setcounter{equation}{0}
\section{Introduction.}
When one wants to quantize theories with constraints it is usually very
important to
first separate them into first and second class constraints  since these
classes of
constraints usually have to be treated in a different manner. However, very
often for
relativistic models it is not possible to do this splitting of the original
constraints in a covariant way. Examples include the superparticle, the
superstring,
p-branes and high rank tensor fields. The advantage of the method we
propose here is
that  no such  explicit splitting of the constraints are required. 
Instead it is
based on the idea to find the maximal involutive set of the constraints
which is a
mixture of first class constraints and  one-half of the second class
ones. We believe that this maximal involutive subset can be covariantly
extracted
from the full set of constraints in many physical models for which there are no
covariant splitting into first and second class constraints. An example is
given at
the end.

Let us first give the setting for our considerations.
We consider a general dynamical theory with finite number of degrees of 
freedom.
(The  generalization to infinite degrees of freedom is straight-forward.)
Given is  a phase space of dimension
$2n$ spanned by the canonical coordinates
$x^I=(q^i, p_i)$,
$i=1,\ldots,n$, with arbitrary Grassmann parities,
$\ve(p_i)=\ve(q^i)=\ve_i$. On this
phase space we have
$m$ constraints
\be
&&T_{\al}(q,p)=0,\quad \al=1,\ldots,m,\quad \ve(T_{\al})\equiv\ve_{\al},
\e{1}
which are not required to be irreducible, since the requirement of covariance may
force us to use dependent constraints.
 We have
\be
&&m=m_1+2m_2+q,
\e{2}
where $m_1$ is the number of independent first class constraints, $2m_2$
the number
of independent second class constraints, and where $q$ is the number of
dependent
constraints. More precisely
\be
&&{\rm rank}\{T_{\al}, T_{\beta}\}\biggr|_{T=0}=2m_2,\quad {\rm
rank}{\dif T_{\al}\over\dif x^I}\biggr|_{T=0}=m_1+2m_2,\nn\\&&
Z^{\al}_{a}T_{\al}=0, \; a=1,\ldots,q,\quad {\rm rank}Z^{\al}_{a}\biggr|_{T=0}=q,
\e{201}
where we have introduced the graded Poisson bracket on the $2n$-dimensional
phase
space.
 There are several different procedures to quantize such a system. There
is
\eg the method of conversion in which one adds new degrees of freedom by
means of
which one may convert the second class constraints into first class ones
which then may
be quantized by the standard procedure for general gauge theories
\cite{BF}. Another method is to remove half of the second class constraints
which in the present case may be stated in the following general
form\footnote{Among other methods for the quantization of theories with
second class constraints there
are the conversion in terms of original variables \cite{LM,BM}, the method
of split
involution \cite{BTL}, and the generalized BRST proposal in \cite{RM}}: Find 
the
maximal involutive subset of
$T_{\al}$. The number of such constraints is $m_1+m_2+s$
where $m_1+m_2$ of them are independent and
where $s\leq q$ is the number of  constraints dependent on the chosen
independent set. The  procedure of quantization which we propose here is
related to
the latter idea but will be formulated in a more invariant way. The basic
ingredient is a projection matrix,
$P_{\al}^{\;\beta}$,
$\ve(P_{\al}^{\;\beta})=\ve_{\al}+\ve_{\beta}$, chosen in such a way that
\be
&&T'_{\al}=P_{\al}^{\;\beta}T_{\beta},
\e{3}
 are constraints  in involution, \ie which satisfy the Poisson algebra
\be
&&\{T'_{\al}, T'_{\beta}\}=U_{\al\beta}^{\prime\;\ga}T'_{\ga}.
\e{4}
$P_{\al}^{\;\beta}$ is also required to be covariant. The idea is to cast
the original
theory  into a general gauge theory where the covariant constraints
$T'_{\al}$ generate the gauge transformations, and where the observables $O$
(including the Hamiltonian) satisfy\footnote{This does not restrict the
generality
of the theory as any observable can be brought to the involution \r{399} by
adding
combinations of the constraints $T_{\al}$ in \r{1}.}
\be
&&\{O,T'_{\al}\}=V_{\al}^{\prime\;\beta}T'_{\beta}.
\e{399}
In this way the quantization problem of the original theory is reduced to
that of
an effective theory with the first class constraints $T'_{\al}$ only.
 In order for this to be possible
$T'_{\al}$ {\em must contain a maximal subset of independent $T_{\al}$ in
involution}.
 More
precisely
$P_{\al}^{\;\beta}$ must  be such that
$T'_{\al}$  contains exactly
$m_1+m_2$ independent constraints and
$m_2+q$ dependent ones some of which may be zero identically. (With respect
to the
original constraints,
$T_{\al}$,  the independent constraints in $T'_{\al}$ contain $m_1$ independent
first class constraints and  $m_2$  of the independent second class
constraints.) We
have
\be
&&{\rm rank}{\dif T'_{\al}\over\dif x^I}\biggr|_{T'=0}=m_1+m_2,
\e{400}
and there exist a function $Z^{\prime\al}_{\al_1}$ with the properties
\be
&&Z^{\prime\al}_{\al_1}T'_{\al}=0,\quad {\rm rank}Z^{\prime\al}_{\al_1}\biggr|_{T'=0}=m_2+q.
\e{401}
Eq.\r{400} suggests the rank condition
\be
&&{\rm rank}P_{\al}^{\;\beta}\biggr|_{T'=0}=m_1+m_2.
\e{402}
Eq.\r{401} may \eg be satisfied by the condition that there should exist a
function $Z^{\prime\al}_{\al_1}$ with the rank $m_2+q$ satisfying
\be
&&Z^{\prime\al}_{\al_1}P_{\al}^{\;\beta}=0.
\e{403}
These properties suggest that $P_{\al}^{\;\beta}$ may be chosen to be a
function satisfying the projection property
\be
&&P_{\al}^{\;\ga}P_{\ga}^{\;\beta}=P_{\al}^{\;\beta},
\e{301}
in which case
\be
&& {\rm
rank}\left(\del_{\al}^{\beta}-P_{\al}^{\;\beta}\right)\biggr|_{T'=0}=m_2+q.
\e{302}

In order to see more explicitly what the condition \r{4} requires we first
notice that
we always may write
\be
&&\{T_{\al}, T_{\beta}\}=C_{\al\beta}+U_{\al\beta}^{\;\;\ga}T_{\ga},
\e{404}
where the separation into the two terms  is purely conventional.
As we shall see our formalism suggests that there always exist  a
particular separation  and a particular choice of $P_{\al}^{\;\beta}$ such that
\be
&&P_{\al}^{\;\ga}C_{\ga\rho}P_{\beta}^{\;\rho}(-1)^{\ve_{\rho}(\ve_{\beta}+1
)}=0.
\e{407}
This expression together with the equation obtained by inserting \r{404}
into
\r{4} yields then the conditions
\be
&&\biggl(P_{\al}^{\;\ga}\{T_{\ga}, P_{\beta}^{\;\rho}\}-
P_{\beta}^{\;\ga}\{T_{\ga}, P_{\al}^{\;\rho}\}(-1)^{\ve_{\al}\ve_{\beta}}+
\{P_{\al}^{\;\rho},
P_{\beta}^{\;\ga}\}T_{\ga}(-1)^{\ve_{\rho}\ve_{\beta}}+\nn\\&&+
P_{\beta}^{\;\eta}P_{\al}^{\;\ga}U_{\ga\eta}^{\;\;\rho}(-1)^{\ve_{\al}(\ve_{
\beta}+\ve_{\eta})}-
U_{\al\beta}^{\prime\;\ga}P_{\ga}^{\;\rho}\biggr)T_{\rho}=0.
\e{406}

\section{Invariant formulation: Classical theory.}
In order
to put the above ideas into a more invariant formulation we have to extend
the phase
space by ghost variables
$\ca^{\al}$, $\bpet_{\al}$ and ghost for ghosts $\ca^{\al_1}$,
$\bpet_{\al_1}$ etc up
to a certain stage $L$ satisfying the properties
\be
&&\ve(\ca^{\al})=\ve(\bpet_{\al})=\ve_{\al}+1, \quad
\ve(\ca^{\al_r})=\ve(\bpet_{\al_r})=\ve_{\al_r}+r+1, \; r=1,\ldots,L, \nn\\
&&\{\ca^{\al},
\bpet_{\beta}\}=\del^{\al}_{\beta},\quad \{\ca^{\al_r},
\bpet_{\beta_r}\}=\del^{\al_r}_{\beta_r}, \; r=1,\ldots,L.
\e{5}
$\ca^{\al}$ and $\bpet_{\al}$ have ghost number one and minus one
respectively, \ie
$gh(\ca^{\al})=1$ and $gh(\bpet_{\al})=-1$, and
$gh(\ca^{\al_r})=r+1$ ($gh(\bpet_{\al_r})=-r-1$). The ghost number $gh(f)$ of a
function $f$ is defined by
\be
&&\{G,f\}=gh(f)f,\quad G\equiv\ca^{\al}\bpet_{\al}(-1)^{\ve_{\al}}
+\sum_{r=1}^L{(r+1)}\ca^{\al_r}\bpet_{\al_r}(-1)^{\ve_{\al_r}+r},
\e{500}
where $G$ is the ghost charge. In terms of the ghost variables \r{5} we
have an odd,
real function
$\Om$ with ghost number one containing the terms
\be
&&\Om=\ca^{\al}T_{\al}+\ca^{\al_1}
Z_{\al_1}^{\;\al}\bpet_{\al}(-1)^{\ve_{\al}}+\sum_{r=2}^L\ca^{\al_r}
Z_{\al_r}^{\;\al_{r-1}}\bpet_{\al_{r-1}}(-1)^{\ve_{\al_{r-1}}}+\nn\\&&
+(-1)^{\ve_{\beta}}\half
\ca^{\beta}\ca^{\al}U_{\al\beta}^{\;\;\ga}\bpet_{\ga}(-1)^{\ve_{\ga}}
+(-1)^{\ve_{\al}}
\ca^{\al}\ca^{\beta_1}U_{\beta_1\al}^{\;\;\al_1}\bpet_{\al_1}
(-1)^{\ve_{\al_1}}+\nn\\&&
+(-1)^{\ve_{\beta}+\ve_{\al}\ve_{\ga}}{1\over6}
\ca^{\ga}\ca^{\beta}\ca^{\al}U_{\al\beta\ga}^{\;\;\al_1}\bpet_{\al_1}
(-1)^{\ve_{\al_1}}+\ldots,
\e{6}
which apart from the first term is a general ansatz. In the first line we
have the
lowest order terms, while the second and third lines explicitly represent
the terms
linear in $\bpet_{\al}$ and $\bpet_{\al_1}$, and the dots mean the
remaining terms
allowed by the conditions $\ve(\Om)=1$, $gh(\Om)=1$.
 All Grassmann parities are determined by $\ve(\Om)=1$. The
functions $Z$ \eg have the Grassmann parity
$\ve(Z_{\al_r}^{\;\al_{r-1}})=\ve_{\al_r}+\ve_{\al_{r-1}}$.
$\Om$ is a BRST-like charge.  However, in the case when
$T_{\al}$ contains second class constraints
$\Om$ may not be required to satisfy the condition $\{\Om, \Om\}=0$ as in the
standard BFV-prescription \cite{BFV}. (The nonzero matrix $C_{\al\beta}$ in
\r{404}
causes the
obstruction.)\footnote{Such a generalized BRST-charge have been used for
second class
constraints for irreducible $T_{\al}$ but without ghost for ghosts in
\cite{RM}.} In
the presence of second class constraints we need therefore a new principle 
which
tells us how to choose the
 terms  in
\r{6} apart from the first one which is a boundary term. We propose here such a
principle by means of which we may also extract a conventional BFV-BRST
charge for
the given theory. This principle requires us first to introduce
 an even, real  function $\Pi$ with ghost number zero given by the ansatz
\be
&&\Pi=\ca^{\al}P_{\al}^{\;\beta}\bpet_{\beta}(-1)^{\ve_{\beta}}+\sum_{r=1}^L
(-1)^{r}
\ca^{\al_{r}}P_{\al_{r}}^{\;\beta_{r}}\bpet_{\beta_{r}}(-1)^{\ve_{\beta_{r}}}
+\ldots,
\e{7}
where the last dots indicates terms containing higher powers in the ghosts. The
matrix functions $P_{\al_r}^{\;\beta_r}$, $r=1,\ldots,L$, have the
Grassmann parity
$\ve(P_{\al_r}^{\;\beta_r})=\ve_{\al_r}+\ve_{\beta_r}$, and the  matrix
function entering in the first term will be the one mentioned in the
introduction,
$\ve(P_{\al}^{\;\beta})=\ve_{\al}+\ve_{\beta}$. The terms in \r{6} and \r{7} 
are
then required to satisfy  the
 following two
invariant conditions
\be
&&\{\Pi,\{\Pi, \Om\}\}=\{\Pi, \Om\},
\e{8}
and
\be
&&\{\Pi,\{\Pi, \{\Om, \Om\}\}\}=\{\Pi, \{\Om, \Om\}\}.
\e{9}
These two conditions imply that
\be
&&\Om'\equiv\{\Pi, \Om\}
\e{10}
satisfies the property
\be
&&\{\Om', \Om'\}=0.
\e{11}
Notice that conditions \r{8} and \r{9} are equivalent to \r{8} and \r{11}.
The odd function $\Om'$ should be a conventional BFV-BRST charge for the
projected
constraints $T'_{\al}$.
$\Om'$ will then  be used in the quantization of  the
original constrained theory.

The reality
conditions of $\Om$ and $\Pi$ may be met by the following choices:
$T_{\al}$,
$\ca^{\al}$, and $\ca^{\al_r}$, $r=1,\ldots,L$, are real, and
$\bpet_{\al}^*=-\bpet_{\al}(-1)^{\ve_{\al}}$,
$\bpet_{\al_r}^*=-\bpet_{\al_r}(-1)^{\ve_{\al_r}+r}$,
$(P_{\al}^{\;\beta})^*=P_{\al}^{\;\beta}(-1)^{\ve_{\beta}(\ve_{\al}+1)}$,
$(P_{\al_r}^{\;\beta_r})^*=
P_{\al_r}^{\;\beta_r}(-1)^{(\ve_{\beta_r}+r)(\ve_{\al_r}+r+1)}$,\\
$(Z_{\al_1}^{\;\al})^*=Z_{\al_1}^{\;\al}(-1)^{(\ve_{\al_1}+1)(\ve_{\al}+1)}$,
$(Z_{\al_r}^{\;\al_{r-1}})^*=Z_{\al_r}^{\;\al_{r-1}}
(-1)^{(\ve_{\al_{r-1}}+r)(\ve_{\al_r}+r)}$, $r=1,\ldots,L$.

Let us to start with  the pure abelian case when  $\{T_{\al},
T_{\beta}\}=C_{\al\beta}$ is a constant and all
 matrix functions, $P$, as well as all $Z$-functions are constants,
 and when $\{T'_{\al},
T'_{\beta}\}=0$. $\Om$ and
$\Pi$ are given by \r{6} and \r{7} up to quadratic terms in the ghosts. In
this case
we find
\be
&&\Om'\equiv\{\Pi, \Om\}=\ca^{\al}T'_{\al}+\ca^{\al_1}
Z_{\al_1}^{\prime\;\al}\bpet_{\al}(-1)^{\ve_{\al}}+\sum_{r=2}^L\ca^{\al_r}
Z_{\al_r}^{\prime\;\al_{r-1}}\bpet_{\al_{r-1}}(-1)^{\ve_{\al_{r-1}}},
\e{110}
where
\be
&&T'_{\al}=P_{\al}^{\;\beta}T_{\beta},
\quad
Z_{\al_r}^{\prime\;\al_{r-1}}=P_{\al_r}^{\;\beta_r}Z_{\beta_r}^{\;\al_{r-1}}
-
Z_{\al_r}^{\;\beta_{r-1}}P_{\beta_{r-1}}^{\;\al_{r-1}},\quad r=1,\ldots,L.
\e{111}
Furthermore, we get
\be
&&\Om''\equiv\{\Pi, \Om'\}=\ca^{\al}T''_{\al}+\ca^{\al_1}
Z_{\al_1}^{\prime\prime\;\al}\bpet_{\al}(-1)^{\ve_{\al}}+
\sum_{r=2}^L\ca^{\al_r}
Z_{\al_r}^{\prime\prime\;\al_{r-1}}\bpet_{\al_{r-1}}(-1)^{\ve_{\al_{r-1}}},\nn\\
\e{112}
where
\be
&&T''_{\al}=P_{\al}^{\;\beta}T'_{\beta},\quad
Z_{\al_r}^{\prime\prime\;\al_{r-1}}=P_{\al_r}^{\;\beta_r}Z_{\beta_r}^{\prime
\;\al_{r-1}}-
Z_{\al_r}^{\prime\;\beta_{r-1}}P_{\beta_{r-1}}^{\;\al_{r-1}},
\quad r=1,\ldots,L.
\e{113}
The condition \r{8}, \ie $\Om''=\Om'$, requires then the  property
\be
&&T''_{\al}=T'_{\al}\quad\Leftrightarrow\quad\biggl(P_{\al}^{\;\ga}P_{\ga}^{
\;\beta}-
P_{\al}^{\;\beta}\biggr)T_{\beta}=0,
\e{1131}
 and
\be
&&Z_{\al_r}^{\prime\prime\;\al_{r-1}}=Z_{\al_r}^{\prime\;\al_{r-1}},\quad
r=1,\ldots,L.
\e{114}
The condition \r{9} or \r{11}, \ie $\{\Om', \Om'\}=0$, requires on the
other hand
\be
&&P_{\al}^{\;\ga}C_{\ga\rho}P_{\beta}^{\;\rho}(-1)^{\ve_{\rho}(\ve_{\beta}+1
)}=0,
\quad
Z_{\al_1}^{\prime\;\beta}T'_{\beta}=0,\quad
Z_{\al_r}^{\prime\;\al_{r-1}}Z_{\al_{r-1}}^{\prime\;\beta_{r-2}}=0,\quad
r=2,\ldots,
L.\nn\\
\e{115}
Since $\Om'$ should be a standard BFV-BRST charge for a reducible theory
\cite{BF2,HT}, we have also the standard rank conditions: For the  ranges 
of the
indices, $\al_r=1,\ldots,k_r$  we have ($\al_0=\al, \;k_0=m_1+2m_2+q$):
\be
&& {\rm
rank}Z^{\prime\al_{r-1}}_{\al_r}\biggr|_{T'=0}=\ga_r, \quad
\ga_r\equiv\sum_{r'=r}^Lk_{r'}(-1)^{r'-r},
\e{116}
where
\be
&&\ga_0={\rm rank}{\dif T'_{\al}\over\dif x^I}\biggr|_{T'=0}=m_1+m_2,
\e{1161}
which serves as a restriction on the ranges $k_r$. Notice that
$\ga_1=m_2+q$. There are several ways in which these conditions may be met by
appropriate choices of the functions in
$\Om$ and $\Pi$. One simple choice is
\be
&&P_{\al_r}^{\;\beta_r}=r\del_{\al_r}^{\beta_r},\;r=2,\ldots,L,
\quad\Rightarrow\quad
Z_{\al_r}^{\prime\prime\;\al_{r-1}}=Z_{\al_r}^{\prime\;\al_{r-1}}=
Z_{\al_r}^{\;\al_{r-1}},\;r=2,\ldots,L,\nn\\
\e{117}
where $Z_{\al_r}^{\;\al_{r-1}}$, $r=2,\ldots,p$, must be chosen to satisfy
\r{115} and
\r{116}.
The condition \r{114} for $r=1$ may then be solved by imposing the
projection property \r{301} on $P_{\al}^{\;\beta}$ in which case we have
\be
&&
Z_{\al_1}^{\prime\prime\;\al}=Z_{\al_1}^{\prime\;\al}=
Z_{\al_1}^{\;\beta}\left(\del_{\al}^{\beta}-P_{\al}^{\;\beta}\right),
\e{118}
which in turn implies that $Z_{\al_1}^{\prime\;\al}$ automatically satisfies the
second condition in \r{115}.

In the case of a general first stage theory ($L=1$) we have
\be
&&\Om'\equiv\{\Pi, \Om\}=\ca^{\al}T'_{\al}+\ca^{\al_1}Z^{\prime
\al}_{\al_1}\bpet_{\al}(-1)^{\ve_{\al}}+\half
\ca^{\beta}\ca^{\al}
U_{\al\beta}^{\prime\;\ga}\bpet_{\ga}(-1)^{\ve_{\beta}+\ve_{\ga}}+\ldots,\nn\\
\e{12}
where $T'_{\al}$ and $Z^{\prime
\al}_{\al_1}$ are given by \r{111} and
\be
&&U_{\al\beta}^{\prime\;\ga}=\{T_{\al}, P_{\beta}^{\;\ga}\}-\{T_{\beta},
P_{\al}^{\;\ga}\}(-1)^{\ve_{\al}\ve_{\beta}}+P_{\al}^{\;\rho}U_{\rho\beta}^{
\;\;\ga}-
P_{\beta}^{\;\rho}U_{\rho\al}^{\;\;\ga}(-1)^{\ve_{\al}\ve_{\beta}}-
U_{\al\beta}^{\;\;\rho}P_{\rho}^{\;\ga}.\nn\\
\e{121}
The condition \r{8}, \ie $\Om''\equiv\{\Pi, \Om'\}=\Om'$ requires \eg
\r{1131} and
\r{114} for
$r=1$, and
\be
&&U_{\al\beta}^{\prime\prime\;\ga}=U_{\al\beta}^{\prime\;\ga}.
\e{122}
Again \r{114} is satisfied
 for the choice \r{117},
$P_{\beta_1}^{\;\al_1}=\del_{\beta_1}^{\al_1}$, and if \r{301} is valid. 
In this
case we have
\be
&&Z^{\prime
\al}_{\al_1}\equiv Z^{\beta}_{\al_1}(\del_{\beta}^{\al}-P_{\beta}^{\;\al}).
\e{13}
 Property \r{301} implies furthermore that
\r{122} requires \r{406} to be satisfied. Thus, $U_{\al\beta}^{\;\;\ga}$ in
$\Om$
may be identified with
$U_{\al\beta}^{\;\;\ga}$ in \r{404} when \r{301} is satisfied. In principle
$U_{\al\beta}^{\prime\;\ga}$ in
\r{12} may contain a term $U_{\al\beta}^{\prime\;\al_1}Z_{\al_1}^{\ga}$
coming from
the following term in $\Pi$:
\be
&&\half
(-1)^{\ve_{\beta}}\ca^{\beta}\ca^{\al}U_{\al\beta}^{\prime\;\al_1}
\bpet_{\al_1}(-1)^{\ve_{\al_1}}.
\e{131}
However, condition \r{122} together with \r{301} excludes such a term.
The nilpotency of
$\Om'$ requires
\r{115}. The second condition in \r{115} is satisfied since \r{403} follows 
when
\r{301} is satisfied.
It is clear that conditions \r{8}
and \r{11}  require the property \r{4} together with all its Jacobi
identities. This
first stage reducible treatment is sufficient if $\al_1=1,\ldots,m_2+r$
since the rank
of
$Z^{\prime
\al}_{\al_1}$ is $m_2+r$.  However, in order for $\al_1$ to be a covariant
index it
might not be possible to satisfy this range condition in which case one is
forced to
consider a higher stage reducible treatment.

In the case when the original
constraints are first class ones, \ie when
\be
&&{\rm rank}\{T_{\al}, T_{\beta}\}\biggr|_{T=0}=0,
\e{14}
it is possible to choose
 $\Om$  to satisfy $\{\Om,\Om\}=0$. $\Om$ is then determined. $\Pi$ may
then be chosen to be the ghost charge $G$ in \r{500} in which case $\Om'=\Om$.
However, it may also be possible to find a $\Pi$ different from $G$
which satisfy
\r{8} and
\r{9}. In this case $\Om'\neq\Om$ but
$\Om'$ will then be canonically equivalent to
$\Om$.

\section{Invariant formulation: Quantum theory.}
At the quantum level all canonical variables above are turned into
operators. We have
then the nonzero fundamental commutation relations (denoting the operators
by the
same symbols as above)
\be
&&[x^i, p_j]=i\hbar\del^i_j,\quad [\ca^{\al},
\bpet_{\beta}]=i\hbar\del^{\al}_{\beta},\quad  [\ca^{\al_1},
\bpet_{\beta_1}]=i\hbar\del^{\al_1}_{\beta_1},\ldots
\e{20}
These operators have the same Grassmann parities as the corresponding classical
variables and all commutators are graded ones.  All real functions in the
classical theory are then turned into hermitian operators which we give in a
Weyl-ordered form below. The ghost charge operator
$G$ is (cf \r{500})
\be
&&G=\half\left(\ca^{\al}\bpet_{\al}(-1)^{\ve_{\al}}-
\bpet_{\al}\ca^{\al}
\right)+\nn\\&&+\sum_{r=1}^L{(r+1)\over
2}\left(\ca^{\al_r}\bpet_{\al_r}(-1)^{\ve_{\al_r}+r}-\bpet_{\al_r}\ca^{\al_r
}\right).
\e{30}
The odd, hermitian
BRST-like charge
$\Om$ with ghost number one is of the form (cf the classical expression \r{6})
\be
&&\Om=\ca^{\al}T_{\al}+\ca^{\al_1}
Z_{\al_1}^{\;\al}\bpet_{\al}(-1)^{\ve_{\al}}+\sum_{r=2}^L\ca^{\al_r}
Z_{\al_r}^{\;\al_{r-1}}\bpet_{\al_{r-1}}(-1)^{\ve_{\al_{r-1}}}+\ldots,\nn\\
&&(i\hbar)^{-1}[G,\Om]=\Om,
\e{21}
where the dots indicate terms of higher powers in the ghosts. The
hermitian projection operator $\Pi$ of ghost number zero is of the form (cf the
classical expression \r{7})
\be
&&\Pi=\half\left(\ca^{\al}P_{\al}^{\;\beta}\bpet_{\beta}(-1)^{\ve_{\beta}}-
\bpet_{\beta}(-1)^{\ve_{\beta}}P_{\al}^{\;\beta}\ca^{\al}(-1)^{\ve_{\al}\ve_
{\beta}}
\right)-\nn\\&&+
\half\sum_{r=1}^L(-1)^r\left(\ca^{\al_r}P_{\al_r}^{\;\beta_r}\bpet_{\beta_r}
(-1)^{\ve_{\beta_r}}
+\bpet_{\beta_r}(-1)^{\ve_{\beta_r}}P_{\al_r}^{\;\beta_r}\ca^{\al_r}
(-1)^{(\ve_{\al_r}+1)(\ve_{\beta_r}+1)}\right)
+\ldots,\nn\\&& (i\hbar)^{-1}[G,\Pi]=0.
\e{22}
The explicit terms in \r{21} and \r{22} are boundary terms chosen in
accordance with
the requirements of the corresponding classical theory. Hermiticity may be
obtained if
we choose
$T_{\al}$,
$\ca^{\al}$, and $\ca^{\al_r}$, $r\geq1$, to be hermitian and
\be
&&\bpet_{\al}\dagg=-\bpet_{\al}(-1)^{\ve_{\al}},\quad
\bpet_{\al_r}\dagg=-\bpet_{\al_r}(-1)^{\ve_{\al_r}+r}, r\geq1,\quad
(P_{\al}^{\;\beta})\dagg=P_{\al}^{\;\beta}
(-1)^{\ve_{\beta}(\ve_{\al}+1)},\nn\\&&
(P_{\al_r}^{\;\beta_r})\dagg=P_{\al_r}^{\;\beta_r}
(-1)^{(\ve_{\beta_r}+r)(\ve_{\al_r}+r+1)}, r\geq1,
\quad
(Z_{\al_1}^{\;\al})\dagg=Z_{\al_1}^{\;\al}(-1)^{(\ve_{\al_1}+1)(\ve_{\al}+1)
},\nn\\&&
(Z_{\al_r}^{\;\al_{r-1}})\dagg=
Z_{\al_r}^{\;\al_{r-1}}(-1)^{(\ve_{\al_r}+r)(\ve_{\al_{r-1}}+r)}, r\geq2.
\e{221}
The dotted terms in \r{21} and \r{22} are required to satisfy  the invariant
conditions
\be
&&(i\hbar)^{-2}[\Pi,[\Pi, \Om]]=(i\hbar)^{-1}[\Pi, \Om],
\e{23}
and
\be
&&(i\hbar)^{-2}[\Pi,[\Pi, [\Om, \Om]]]=(i\hbar)^{-1}[\Pi, [\Om, \Om]],
\e{24}
which implies that
\be
&&\Om'\equiv(i\hbar)^{-1}[\Pi, \Om]
\e{25}
satisfies the property
\be
&&[\Om', \Om']=0.
\e{26}
The hermitian operator $\Om'$ has ghost number one since
\be
&&(i\hbar)^{-1}[G,\Om]=\Om,\quad (i\hbar)^{-1}[G, \Pi]=0\quad\Rightarrow\quad
(i\hbar)^{-1}[G,
\Om']=\Om'.
\e{31}
It is also required to be a BFV-BRST charge operator to be used in the
conventional
way.
We expect that there always exist solutions to all these conditions
when we have finite number of degrees of freedom, at least if we relax the
requirement
of covariance which might lead to obstructions. The reason is that  we may
solve the
conditions for the abelian case treated classically in the previous
section, and that
we expect that the general case may be obtained by a unitary transformation
of this
abelian case at least locally. The latter property is true in the standard BFV
treatment \cite{BF3}.

One may notice that if we choose $\Pi=G$ then  \r{23} is satisfied by
construction
but
\r{24} is only satisfied if
$\Om$ is nilpotent in which case $\Om'=\Om$. This is possible only if the
original
constraints are purely first class ones. In the latter case we also obtain
an $\Om'$
which is unitary equivalent to $\Om$ if there exists a $\Pi\neq G$.

We need also a BRST invariant Hamiltonian. The original Hamiltonian may
always be
turned into a Hamiltonian $\cH_0$ satisfying the observability condition
\r{399} by
simply adding linear combinations of the constraints $T_{\al}$, \ie $\{\cH_0,
T'_{\al}\}=V_{\al}^{\;\beta}T'_{\al}$. We have then after quantization
($(V_{\al}^{\;\beta})^{\dag}=V_{\al}^{\;\beta}(-1)^{\ve_{\beta}(\ve_\al+1)}$)
\be
&&(i\hbar)^{-1}[\cH, \Om']=0,\quad
\cH=\cH_0+\half\left(\ca^{\al}V_{\al}^{\;\beta}\bpet_{\beta}(-1)^{\ve_{\beta}}-
\bpet_{\beta}(-1)^{\ve_{\beta}}V_{\al}^{\;\beta}\ca^{\al}(-1)^{\ve_{\al}\ve_
{\beta}}
\right)+\ldots.\nn\\
\e{42}
Due to the condition \r{23} this implies $[\cH', \Om']=0$ where
\be
&&\cH'\equiv(i\hbar)^{-1}[\Pi, \cH]=\half\left(\ca^{\al}(P_{\al}^{\;\beta}
V_{\beta}^{\;\rho}-
V_{\al}^{\;\beta}P_{\beta}^{\;\rho})\bpet_{\rho}(-1)^{\ve_{\rho}}-\right.\nn
\\&&\left.\quad
\quad
\quad \quad \quad \quad
-\bpet_{\rho}(-1)^{\ve_{\rho}}(V_{\beta}^{\;\rho}P_{\al}^{\;\beta}-
P_{\beta}^{\;\rho}V_{\al}^{\;\beta})\ca^{\al}(-1)^{\ve_{\beta}(\ve_\rho+\ve_
\al+1)}
\right)+\ldots.
\e{43}
We expect that $\cH'=(i\hbar)^{-1}[\rho, \Om']$. Notice in this connection that
$\cH\rightarrow
\cH+(i\hbar)^{-1}[\psi,\Om']$ implies $\cH'\rightarrow \cH'+(i\hbar)^{-1}[\psi+\psi',\Om']$ where
$\psi'\equiv(i\hbar)^{-1}[\Pi,
\psi]$.

$\Om'$ constitutes the BRST charge in the minimal sector for the original
theory. A
complete BRST quantization requires  the introduction of antighosts 
and Lagrange
multipliers and their conjugate momenta up to a certain extention required
by the
the prescription given in \cite{BF2}. The total BRST charge $Q'$ has then
the form
\be
&&Q'=\Om'+\sum_{s'=0}^L\sum_{s=s'}^L\pi_s^{s'}\pet_s^{s'},
\e{44}
and the total Hamiltonian $H$ is
\be
&&H=\cH+(i\hbar)^{-1}[\Psi, Q'],\quad (i\hbar)^{-1}[H, Q']=0,
\e{45}
where
\be
&&\Psi=\sum_{s'=0}^L\sum_{s=s'}^L\left(\bac^{s'}_s\chi_s^{s'}+
\bar{\chi}_s^{s'}\la_s^{s'}\right)
\e{46}
is an odd gauge-fixing fermion.
In \r{44} and \r{46} we have used a short-hand notation which may be
understood by a
comparison with
\cite{BF2}. $\bac^{s'}_s$ and $\la_s^{s'}$ represents antighosts and
Lagrange-multipliers (extra ghosts), and
$\pet_s^{s'}$ and $\pi_s^{s'}$ their conjugate momenta. $\chi_s^{s'}$ and
$\bar{\chi}_s^{s'}$ are gauge fixing functions (see
\cite{BF2}). $Q'$ is always nilpotent when $\Om'$ is nilpotent. Physical
states are
determined by the condition
\be
&&Q'|phys\hb=0.
\e{47}
It is clear that there also exist an extended $\Om$ and an extended projection
operator $\Pi$, denoted $Q$ and $\tilde{\Pi}$ respectively, satisfying
\be
&&Q''\equiv(i\hbar)^{-1}[\tilde{\Pi}, Q']=Q', \quad Q'\equiv(i\hbar)^{-1}[\tilde{\Pi}, Q].
\e{48}
The last property implies
\be
&&\tilde{\Pi}|phys\hb=|phys\hb'.
\e{49}
Since $\tilde{\Pi}$ contains a ghost dependence which is close to an
extended ghost
charge it seems  as if we may impose the
condition
\be
&&\tilde{\Pi}|phys\hb=0
\e{50}
without affecting the true physical degrees of freedom that  contribute to the
BRST cohomology. Maybe it is even possible to impose $Q|phys\hb=0$. However, in
this case the BRST invariant Hamiltonian $H$ satisfying \r{45} is required to
satisfy the stronger conditions $[H,Q]=0$ and $[H,\tilde{\Pi}]=0$.

\section{An example}
Our method may \eg be applied to the Ferber-Shirafuji "twistorized"
particle model
\cite{AF,TS}. Its Hamiltonian constraint analysis was given in \cite{BMRS},
where
the separation into first and second class constraints only was made in a
translation
noninvariant way. The simplest model of this type is the $d=4$ twistor 
model for
the massless particle. (For a detailed review of the twistorized models, see
\cite{Sor}.) The action is
\be
&&S=\half\int d\tau \sigma_{\mu a \da} \dx^{\mu}\la^a\bla^{\da},
\e{32}
where $\mu=0,1,2,3;$ $a,\da=1,2$. $\sigma_{\mu a \da}$ is a Pauli matrix, and
$\la$, $\bla$ are bosonic SL(2,C) spinors. The indices $a, \da$ are raised and
lowered by $\ve^{ab}$, $\ve_{ab}$ ($\ve^{12}=\ve_{21}=1$). The complete set
of independent constraints is
\be
&&p_{\mu}-\sigma_{\mu a \da}\la^a\bla^{\da}=0,\quad p_a=0,\quad
\bp_{\da}=0,
\e{33} where $p_{\mu}$, $p_a$, and $\bp_{\da}$ are conjugate momenta to
$x^{\mu}$,
$\la^a$, and
$\bla^{\da}$. This set  contains two first class constraints and six
second class ones. The explicit separation given in \cite{BMRS} involved the
combinations
$\mu_a=\sigma_{\mu a \da}\bla^{\da} x^{\mu}$ and $\bmu_{\da}=\sigma_{\mu a
\da}\la^a x^{\mu}$ which violate manifest translation invariance. According
to our
approach we have not to separate the constraints into first and second
class ones,
but to find a set of  constraints {\em in involution} containing the maximal
independent set which here has {\em five elements}. There are two covariant
choices  to pick five such constraints. The first choice is
\be
&&p_{\mu}-\sigma_{\mu a \da}\la^a\bla^{\da}=0, \quad i(\la^a
p_a-\bla^{\da}\bp_{\da})=0.
\e{34}
The second option is
\be
&&\bp_{\da}=0, \quad p_a=0, \quad p^{\mu}p_{\mu}=0.
\e{35}
It seems very difficult to find a projection matrix $P_{\al}^{\;\beta}$
which takes
us from the set \r{33} to the set \r{34} or \r{35} in a covariant manner.
Even if it
would be possible we would have to deal with a higher reducible situation
due to the
difficulty to find a covariant $Z^{\al}_{\al_1}$ with $\al_1=1,2,3$.
However, if we
from the very beginning consider the following reducible set
\be
&&T_{\al}\equiv\biggl( p_{\mu}-\sigma_{\mu a \da}\la^a\bla^{\da}, \;
\bp_{\da},
\; p_a, \; p^{\mu}p_{\mu}, \; i(\la^a
p_a-\bla^{\da}\bp_{\da})\biggr),
\e{36}
which constitutes ten covariant constraints out of which only eight are
independent,
then it is possible to project out the subset \r{34} or \r{35} in a trivial
manner
with a covariant projection matrix
$P_{\al}^{\;\beta}$. Furthermore, the resulting theory will be a first stage
reducible theory, and both $\Om$ and $\Pi$ and thereby $\Om'$ will be 
manifestly
invariant. Notice that $T_{\al}=0$, where $T_{\al}$ is given by  \r{36},  is
completely equivalent to
\r{33}.

\noindent
{\bf Acknowledgements}:

I.A.B. and S.L.L. would like to thank Lars Brink for
his very warm hospitality at the
Department of Theoretical Physics, Chalmers
and G\"oteborg University.
 The work of I.A.B. is supported by the grants 99-01-00980 and
99-02-17916 from Russian foundation for basic researches (RFBR) and by the
President
grant 00-15-96566 for supporting leading scientific schools. S.L.L. is
supported by
STINT guest research stipend, and in parts by grants 00-02-17956 from RFBR and
E00-33-184 from Russian Ministry of Education. The work is partially
supported by INTAS grant 00-00262.  S.L.L. is grateful to I.Bandos and
D.Sorokin for
some references and discussions on the twistor particle models.

\end{document}